\documentclass[a4paper,twocolumn,superscriptaddress,11pt]{quantumarticle}

\usepackage[utf8]{inputenc}  
\usepackage[T1]{fontenc}     
\usepackage[british]{babel}  
\usepackage[sc,osf]{mathpazo}
\usepackage[scaled=0.86]{berasans}  
\usepackage[scaled=1.03]{inconsolata} 
\usepackage[usenames,dvipsnames]{color} 
\usepackage[x11names]{xcolor} 
\usepackage[colorlinks,citecolor=violet,linkcolor=SpringGreen4,urlcolor=blue]{hyperref}  
\usepackage{graphicx} 
\usepackage[babel]{microtype}  
\usepackage{amsmath,amssymb,amsthm,bm,amsfonts,mathrsfs,bbm} 
\usepackage{xspace}  
\usepackage{pgfplots}  
\usepackage{cite}

\DeclareMathOperator{\Tr}{Tr}
\DeclareMathOperator{\sign}{sgn}

\newcommand{\bra}[1]{\left\langle #1 \right|}
\newcommand{\ket}[1]{\left| #1 \right\rangle}

\newcommand{\ba}{\begin{eqnarray}}
\newcommand{\ea}{\end{eqnarray}}
\newcommand{\ban}{\begin{eqnarray*}}
\newcommand{\ean}{\end{eqnarray*}}
\newcommand{\ave}[1]{\langle#1\rangle}

\begin{document}
\title{Better local hidden variable models for two-qubit Werner states and an upper bound on the Grothendieck constant $K_G(3)$}

\author{Flavien Hirsch}
\affiliation{Groupe de Physique Appliqu\'ee, Universit\'e de Gen\`eve, CH-1211 Gen\`eve, Switzerland}

\author{Marco T\'ulio Quintino}
\affiliation{Groupe de Physique Appliqu\'ee, Universit\'e de Gen\`eve, CH-1211 Gen\`eve, Switzerland}
\affiliation{Department of Physics, Graduate School of Science, The University of Tokyo, 7-3-1 Hongo, Bunkyo-ku, Tokyo, Japan}

\author{Tam\'as V\'ertesi}
\affiliation{Institute for Nuclear Research, Hungarian Academy of Sciences,
H-4001 Debrecen, P.O. Box 51, Hungary}

\author{Miguel Navascu\'es}
\affiliation{Institute for Quantum Optics and Quantum Information (IQOQI), Boltzmangasse 3, 1090 Vienna, Austria}

\author{Nicolas Brunner}
\affiliation{Groupe de Physique Appliqu\'ee, Universit\'e de Gen\`eve, CH-1211 Gen\`eve, Switzerland}

\date{\today}  

\begin{abstract}
We consider the problem of reproducing the correlations obtained by arbitrary local projective measurements on the two-qubit Werner state $\rho = v \ket{\psi_-} \bra{\psi_-} + (1- v )\openone/4$ via a local hidden variable (LHV) model, where $\ket{\psi_-}$ denotes the singlet state. We show analytically that these correlations are local for $ v = 999\times689\times{10^{-6}}$ $\cos^4(\pi/50) \simeq  0.6829$. In turn, as this problem is closely related to a purely mathematical one formulated by Grothendieck, our result implies a new bound on the Grothendieck constant $K_G(3) \leq 1/v \simeq 1.4644$. We also present a LHV model for reproducing the statistics of arbitrary POVMs on the Werner state for $v \simeq 0.4553$. The techniques we develop can be adapted to construct LHV models for other entangled states, as well as bounding other Grothendieck constants.

\end{abstract}

\maketitle

A quantum Bell experiment consists of two (or more) distant observers performing local measurements on a shared entangled quantum state. Remarkably the predictions of quantum theory are here incompatible with a natural definition of locality formulated by Bell \cite{bell64}. Specifically, the statistics of certain quantum Bell experiments are found to be nonlocal (in the sense of Bell), as witnessed via violation of Bell inequalities. This phenomenon, referred to as quantum nonlocality, represents a fundamental aspect of quantum theory as well as a central resource for quantum information processing \cite{brunner_review}.

Understanding the exact relation between entanglement and quantum nonlocality is a central problem in the foundations of quantum theory, with implications for quantum information processing. While the use of an entangled state is necessary for observing quantum nonlocal correlations, it is interesting to ask if the converse link also holds. That is, can any entangled state lead to a Bell inequality violation, when performing a set of (judiciously chosen, and possibly infinitely many) local measurements? For pure entangled states, the answer turns out to be positive \cite{gisin91}. For mixed entangled states, the situation is more complex, as first discovered by Werner \cite{werner89}, who presented a class of entangled quantum states (now referred to as Werner states) which admit a local hidden variable (LHV) model for any possible local projective measurements. Therefore such states, while being entangled---i.e. inseparable at the level of the Hilbert space---can never lead to nonlocal correlations. Notably, while Werner's original model focused on projective measurements, Barrett \cite{barrett02} presented a LHV model considering the most general non-sequential measurements, i.e. POVMs. These early results triggered much interest, and subsequent works presented various classes of entangled states admitting LHV models \cite{almeida07,wiseman07,hirsch13,bowles14,sania14,bowles15,bowles15b,nguyen16}, including results for the multipartite case \cite{toth05,bowles16,augusiak14}; see \cite{augusiak_review} for a recent review. More sophisticated Bell scenarios have also been explored\footnote{For instance, one can consider a Bell test in which each observer can perform a sequence of measurements, see e.g. S. Popescu, Phys. Rev. Lett. {\bfseries 74}, 2619 (1995). Interestingly nonlocality can be activated in these more complex Bell scenarios, that is, some local states can turn out to violate a Bell inequality after a suitable sequence of measurements.}, but will not be discussed here.

In parallel, Ac\'in, Gisin and Toner \cite{acin06}, based on previous work by Tsirelson \cite{tsirelson93}, established a direct connection between these questions and a purely mathematical problem discussed by Grothendieck \cite{grothendieck}. In particular, the problem of determining the range of visibilities $v \leq v_c$ ($v_c $ denoting the critical visibility) for which the two-qubit Werner state

\ba \label{werner}
\rho(v) = v \ket{\psi_-} \bra{\psi_-} + (1- v ) \frac{\openone}{4},
\ea
where $\ket{ \psi_-}=(\ket{01}-\ket{10})/\sqrt{2}$ is the singlet state and $\frac{\openone}{4}$ is the maximally mixed state, admits a LHV model for arbitrary projective measurements, is directly related to the Grothendieck constant (of order 3) $K_G(3)$. Specifically, one has that $v_c = 1 / K_G(3)$. While the exact value of the Grothendieck constants are generally unknown, existing (upper) bounds can be used for deriving lower bounds on $v_c$. Notably, a result of Krivine \cite{krivine} for bounding $K_G(3)$ implies that $v_c \geq 0.6595$. Note also that upper bounds on $v_c$ can be obtained by demonstrating explicit Bell inequality violations. From the well-known Clauser-Horne-Shimony-Holt Bell inequality, it follows that $v_c \leq 1/ \sqrt{2} \simeq 0.7071$. This was later improved to $v_c \leq 0.7054$ \cite{vertesi_gro,bobo15} and very recently to $v_c \leq 0.7012$ \cite{brierley16}.

More generally, Grothendieck's problem has implications in many areas of mathematics. It had first major impact on Banach space theory and in $\mathcal{C}^*$-algebra theory. More recently, it impacted graph theory and computer science. For more details, we refer to the following review \cite{pisier15}.

In this work, we present better LHV models for two-qubit Werner states. Let us set 
\begin{equation} 
\label{v1}
v_1 = \frac{999 \cdot 689}{10^6}\cos^4\left(\frac{\pi}{50}\right).
\end{equation}
We first consider the case of projective measurements and prove analytically that
\ba \label{main}
 v_c \geq v_1  \simeq  0.6829 \Leftrightarrow K_G(3) \leq \frac{1}{v_1} \simeq 1.4644 \quad
\ea
which also leads to a better bound on the Grothendieck constant, as stated above. This result is derived by combining two recently introduced numerical methods. The first one is an algorithmic method for constructing LHV models \cite{hirsch15,cavalcanti15}, which is applicable to arbitrary entangled states. The second is a numerical optimization algorithm for estimating the distance between a point and a convex set in $\mathbb{R}^d$ \cite{brierley16}. From the output of these numerical methods, we construct the LHV model analytically.

Our second result is a better LHV model for the two-qubit Werner state considering arbitrary POVMs. The model works for  $v \leq 2v_1/3 \simeq 0.4553$.  The proof relies on the fact that the statistics of arbitrary POVMs featuring a certain level of noise can be reproduced exactly via projective measurements only. By applying this observation to the LHV model we construct for projective measurements, the result follows.

More generally, we believe that the methods presented here open promising new possibilities for the construction of LHV models for quantum states, as well as for other convex set membership problems, such as steerability of quantum states (see e.g. references~\cite{wiseman07,sania14}). We conclude by discussing several possible directions for future research.

\section{Concepts and notations}

Consider an experiment in which two distant parties share a two-qubit Werner state \eqref{werner}, perform some local measurements labelled by $x,y$ and obtain outputs $a,b$, respectively. The statistics of the experiment is given by 
\ba  \label{probQ}
p(ab|xy) = \Tr[\rho(v) A_{a|x} \otimes B_{b|y}].
\ea
Here $A_{a|x}$ and $B_{b|y}$ are the operators representing the local measurements of Alice and Bob. They satisfy positivity and normalization, i.e. $A_{a|x} \geq 0$ and $\sum_a  A_{a|x} = \openone $, and similarly for $B_{b|y}$. These represent general POVMs, which we will consider in the second part of the paper. In the first part, however, we focus on the case of local projective measurements, i.e. adding the constraints $a,b \in \{ -1, +1\}$ and $A_{a|x}A_{a'|x} = \delta_{aa'} A_{a|x}$ for all $x,a,a'$ and similarly for $B_{b|y}$.

Our goal is to determine the range of visibilities $v$ for which the Werner state \eqref{werner} admits a LHV model. That is, the measurement statistics \eqref{probQ} can be decomposed as
\ba
p(ab|xy) = \int  q(\lambda) p_A(a|x \lambda) p_B(b|y \lambda)  d \lambda
\ea
where $\lambda$ represents the local variable, distributed according to the density $q(\lambda)$, and the distributions $p_A(a|x \lambda)$ and $p_B(b|y \lambda)$ are Alice and Bob's local response functions. If such a decomposition can be found for all local projective measurements, the state $\rho(v)$ is said to be local for projective measurements. Moreover, if this decomposition can be extended to all local POVMs, $\rho(v)$ is termed local for POVMs.

As mentioned above the case of projective measurements has a strong connection to the Grothendieck constant, a mathematical constant arising in the context of Banach space theory. Local dichotomic projective qubit measurements are conveniently described via observables of the form

\ba 
O_{\hat{x}} =  \hat{x} \cdot \vec{\sigma}  \; , \;  O_{\hat{y}} =  \hat{y} \cdot \vec{\sigma} 
\ea
where the measurement directions are given by unit vectors $\hat{x}$ and $\hat{y}$ on the Bloch sphere, i.e.  $\hat{x},\hat{y} \in \mathbb{R}^3$, and $\vec\sigma=(\sigma_x,\sigma_y,\sigma_z)$ is the vector of Pauli matrices. The measurement statistics of the Werner state \eqref{werner} are then simply characterized by the expectation values
\begin{align} \label{corrs}
\nonumber  \ave{a} &= \Tr[ O_{\hat{x}} \otimes \openone \rho(v)] = 0   \\
 \nonumber \ave{b}  &= \Tr[ \openone \otimes O_{\hat{y}}  \rho(v)] = 0 \\
 \ave{ab} &= \Tr[ O_{\hat{x}} \otimes O_{\hat{y}}  \rho(v) ]  = - v \, (\hat{x} \cdot \hat{y}).
\end{align}
The problem is now to find the largest visibility, $v_c$, such that the above statistics admits a LHV model. For any visibility $v>v_c$, Bell inequality violation is then possible, even though the Bell test may require an infinite number of local measurements.

Interestingly this problem can be directly related to another purely abstract problem, discussed by Grothendieck. Consider any possible $m \times m$ matrix $M$ such that
\ba \label{M_L}
\left| \sum_{i,j=1}^m M_{ij} \alpha_i \beta_j   \right| \leq 1
\ea
for any real numbers $\alpha_i, \beta_j \in [ -1, +1 ] $. Then, $K_G(n)$ is the smallest number such that
\ba \label{M_Q}
\left| \sum_{i,j=1}^m M_{ij} \hat{\alpha}_i \cdot \hat{\beta}_j   \right| \leq K_G(n)
\ea
for any unit vectors $\hat{\alpha}_i , \hat{\beta}_j  \in \mathbb{R}^n$ and for any matrix $M$.
This defines a set of numbers $K_G(n)$, called the Grothendieck constants of order $n$. The Grothendieck constant is then defined as $K_G = \lim_{n \rightarrow \infty} K_G(n)$. While the exact values of these constants is not known in general (except for $n=2$ where $K_G(2)= \sqrt{2}$), lower and upper bounds were proven, see e.g. \cite{finch,bra11}. Of particular relevance to the present work is the constant of order 3, which relates to $v_c$. Indeed one has that
\ba \label{connection} v_c = \frac{1}{K_G(3)},
\ea as shown in Ref. \cite{acin06}, see Theorem 1. This connection follows from early work by Tsirelson \cite{tsirelson93}, who connected the Grothendieck's problem to Bell inequalities. Basically, the matrix $M$ is associated to a Bell inequality, for which the local bound is 1, see equation \eqref{M_L}. The largest possible violation of this Bell inequality requires, in general, a maximally entangled state of dimension $d \times d$, where $d= 2^{ \lfloor \frac{n}{2} \rfloor}$ \cite{tsirelson93}. It then follows that $K_G(3)$ is the largest possible Bell violation for a maximally entangled two-qubit state, from which \eqref{connection} follows. We refer the reader to Ref. \cite{acin06} for more details.

\section{LHV model for projective measurements}

Our main result is the construction of a LHV model for projective measurements on Werner states $\rho(v)$ for a visibility $v_1$ in equation~(\ref{v1}), whose approximate value is $v_1\simeq 0.6829$. This implies the novel bounds on the critical visibility $v_c$ and hence also on the Grothendieck constant $K_G(3)$ stated in equation \eqref{main}.

We make use of a recently developed method for constructing LHV models for entangled quantum states \cite{hirsch15,cavalcanti15}. The method is algorithmic, in the sense that it is applicable to any entangled state in principle, and can be efficiently implemented on a standard computer (at least for low dimensions). For the case of interest to us, i.e. local projective qubit measurements on a Werner state, the method can be intuitively explained as follows.

Consider a finite set of qubit projective measurements, represented by a finite set of Bloch vectors $\hat{u}_i$, with $i=1,...,m$. These vectors form a polyhedron $\mathcal{P}$ contained in the Bloch sphere. Typically, the vectors will be chosen rather uniformly over the sphere, such that the radius of the largest sphere inscribed in $\mathcal{P}$ (and centered at the origin) is close to 1. We refer to this radius as the `shrinking factor' $\eta$ of the polyhedron $\mathcal{P}$.

Next we consider the measurement statistics obtained by performing the above set of local measurements (for both Alice and Bob) on the Werner state. Specifically, we get
\ba \label{stat}
p(ab|xy) = \Tr[\rho(v) A_{a|x} \otimes B_{b|y}]
\ea
with $A_{a|x} = \frac{1}{2} (\openone + a \, \hat{u}_x \cdot \vec{\sigma})$ and $B_{b|y} =  \frac{1}{2} (\openone + b \, \hat{u}_y \cdot \vec{\sigma})$, where $a,b\in\{\pm1\}$. Since we consider a finite set of $m$ measurement settings (for both Alice and Bob), one can find the maximal visibility $v^*$ for which the above measurement statistics admits a LHV model. In practice, this can be done efficiently (at least for $m \leq 10$) using linear programming, see e.g. \cite{brunner_review}.

As the Werner state $\rho(v^*)$ is local for the set of measurements given by Bloch vectors $\hat{u}_i$, it follows that $\rho(v^*)$ is also local for any noisy measurement of the form
\ba \label{noisy}
A^\eta_{a|x} &=& \frac{1}{2} (\openone + a \,  \eta \hat{u}_x \cdot \vec{\sigma})\nonumber \\
B^\eta_{b|y} &=& \frac{1}{2} (\openone + b \,  \eta \hat{u}_y \cdot \vec{\sigma}).
\ea
Note that the above measurements form a continuous set, forming a `shrunk' Block sphere, given by vectors $ \eta \hat{u}$ where $\eta$ is the shrinking factor of the polyhedron $\mathcal{P}$. As these shrunk vectors lie on the sphere inscribed in $\mathcal{P}$, they can be expressed as convex combinations of the finite set of vectors $ \hat{u}_i$. One can then show that the distribution $p(ab|\hat{x} \hat{y}) = \Tr[\rho(v^*) A^\eta_{a|\hat{x}} \otimes B^\eta_{b|\hat{y}}]$ is local, for any possible measurement directions $\hat{x} $ and $\hat{y}$. This follows from the linearity of the trace rule, and from the fact that the noisy measurement operators $A^\eta_{a|x}$, respectively $B^\eta_{b|y}$, can be expressed as convex combinations of the noiseless operators $A_{a|x} $, respectively $B_{b|y}$; see  \cite{hirsch15,cavalcanti15} for more details.

Finally, note that the statistics of the noisy measurements \eqref{noisy} on $\rho(v^*)$ are in fact equivalent to the statistics of noiseless measurements on a slightly more noisy Werner state $\rho(\eta^2v^*)$. Indeed, it is straightforward to verify that
\ba
\Tr[\rho(v^*) A^\eta_{a|\hat{x}} \otimes B^\eta_{b|\hat{y}}] = \Tr[\rho(\eta^2 v^*) A_{a|\hat{x}} \otimes B_{b|\hat{y}}] \nonumber
\ea
which again holds for any possible measurement directions $\hat{x} $ and $\hat{y}$. We thus conclude that the Werner state $\rho(\eta^2v^*)$ admits a LHV model for all local projective measurements.

As mentioned above, this method can be implemented in practice using standard tools when considering sets of relatively few measurements ($m \leq 10$). In Ref. \cite{hirsch15}, this was used to construct a LHV model for the Werner state for visibilities up to $v = 0.54$. While this construction improves on Werner's original model, which attained $v=1/2$, it does not reach the best known value so far of $v=0.6595$ obtained in Ref. \cite{acin06} based on the connection to the Grothendieck constant. However, a remarkable feature of the above algorithm is the fact that it will converge to $v_c$ when $m \rightarrow \infty$ \cite{hirsch15}. Therefore by running the method for finite sets of measurements featuring a large (but nevertheless finite) number of vectors, one can expect to approach the optimal visibility $v_c$. In particular, one may expect to overcome the best known value of $v=0.6595$ in case the latter is suboptimal. This is precisely what we implemented, using sets containing up to $m=625$ vectors. This allows us to obtain the new bounds stated in equation \eqref{main}.

It is however non-trivial to run the algorithm for sets containing so many vectors. Let us discuss why. Since, the local marginals vanish (see Eq.~(\ref{corrs})), we can restrict ourselves to the set of joint correlation terms $\{\ave{a_x b_y}\}_{x,y}$. Therefore, in case of $m$ binary measurements per party, the local polytope with completely random marginals, which we call correlation polytope and denote by $\mathcal{L}$, lives in dimension $m^2$. Each vertex of $\mathcal{L}$ corresponds to a local deterministic strategy $\lambda=(a_1,a_2,\ldots,a_m,b_1,b_2,\ldots,b_m)$, where each $a_x,b_y$, $(x,y=1,\ldots,m)$, may take the values $\pm 1$, which amounts to $2^{2m}$ distinct strategies. A given $\lambda$ strategy translates to a vertex $\vec D_{\lambda}$, which is a $m\times m$ matrix with entries $D_{\lambda}(x,y)=a_x b_y$. Hence, the polytope $\mathcal{L}$ features altogether $2^{2m}$ vertices.

Our goal is now to decompose a given quantum point $\vec q$, whose entries are $q(x,y) = \ave{a_x b_y} = -v\hat x\cdot\hat y$, as a convex combination of deterministic vertices: $\vec q = \sum_{\lambda}w_{\lambda}\vec D_{\lambda}$. This proves that $\vec q$ is local. In principle, this problem can be solved via linear programming. However, for $m=625$ settings, even \emph{inputting} all deterministic strategies to our linear programming solver is completely out of reach.

In order to circumvent these problems, we resort to a modified version of Gilbert's algorithm \cite{gilbert96}, a popular collision detection method used for instance by the video game industry. This algorithm can provide a $\vec q_{\epsilon}$ such that $\|\vec q - \vec q_{\epsilon}\|\le \epsilon$ without a full vertex characterization of the local polytope. The reader is referred to Ref.~\cite{brierley16} for more details about this algorithm and its extension including convergence properties and further applications in quantum information. The algorithm is iterative and is given by:

\begin{enumerate}
\item Set $i=0$ and pick an arbitrary point $\vec q_i$ inside the polytope $\mathcal{L}$.
\item Given the point $\vec q_i$ and the target point $\vec q$, run an oracle which maximizes the overlap $(\vec q-\vec q_i)\cdot\vec l$ over all $\vec l\in\mathcal{L}$. Let us denote the local point $\vec l$ returned by the oracle by $\vec l_i$.
\item Find the convex combination $\vec q_{i+1}$ of $\vec q_i$ and $\vec l_i$ that minimizes the distance $\|\vec q - \vec q_{i+1}\|$.
\item Let $i=i+1$ and go to Step 2 until the distance $\|\vec q - \vec q_{i}\|\le\epsilon$. Return $\vec q_{\epsilon}\equiv q_{i}$.
\end{enumerate}

Note that the distance $\|\vec q - \vec q_{i}\|$ is a decreasing function of $i$, and actually, when $\vec q$ happens to lie inside $\mathcal{L}$, the algorithm is guaranteed to stop after a number of steps $O(1/\epsilon^2)$~\cite{gilbert96}. Since maximizing the overlap $(\vec q-\vec q_i)\cdot\vec l$ over all local vectors $\vec{l}$ is a NP-hard problem, in Step 2 we must make use of a heuristic method, described in Appendix~A.

\emph{Analytic lower bound for $v_c$.} We now discuss explicitly the procedure we implemented in order to obtain the new bound  \eqref{main} on $v_c=1/K_G(3)$. It is important to note that, while our procedure is based on implementing on a computer the above methods (hence giving a numerical result), the final result is proven analytically. This is done as follows.

The finite set of measurement settings we use is based on a family of polyhedra parameterized by an integer $n$ which results in $m = n^2$ ($m=n^2-n+1$) vertices in case of $n$ odd (even). The shrinking factor of this polyhedron is given by $\eta = \cos^2(\pi/2n)$. Both the construction of the polyhedra in terms of unit vectors $\hat{u}(i_1,i_2)$ and the proof regarding the value of the shrinking factor can be found in Appendix~B. We could implement the calculation up to $n=25$ (i.e. $m=n^2=625$ settings) for which we find the lower bound \eqref{main}.

We set the initial visibility of the Werner state to $v_0=689/1000$, which is guessed to be close to $v_c$. Combined with the above $m=625$ measurement settings (for both Alice and Bob), we obtain the target quantum point $\vec q$.

After $23 \times 10^6$ iterations of the algorithm (\textit{i.e.} we repeat $23 \times 10^6$ times the steps 2-4 of the algorithm),  which was completed on a standard desktop PC within a week, we get numerically a point $\vec q_{\epsilon}$ such that $\|\vec q - \vec q_{\epsilon}\|\le 9.8484\times 10^{-6}$. We then truncate $\vec q_{\epsilon}$ up to $k=16$ digits, which results in the rational point $\vec q_r$. Note that $\vec q_r$ is now local by construction; see Appendix C. We now have that
\begin{equation}
\vec q = \vec q_r + \vec q_\textrm{junk},
\end{equation}
where $\vec q_\textrm{junk}$ takes care of the (small) difference between the analytical points $\vec q$ and $\vec q_r$.
Let us now slightly shrink the point $\vec q$ towards the centre of the local correlation polytope (the origin) by rescaling $\vec q$ with a factor $\nu$ close to (but strictly smaller than) 1:
\begin{equation}
\label{vecpq2}
\nu\vec q = \nu\vec q_r + (1-\nu)\frac{\nu\vec q_\textrm{junk}}{1-\nu}.
\end{equation}
Clearly, we see that $\nu\vec q$ is provenly local if the point $\vec x = \nu\vec q_\textrm{junk}/(1-\nu)$ is local as well. By rearranging \eqref{vecpq2}, we have that
\begin{equation}
\label{defx}
\vec x = \frac{\nu}{1-\nu}\left(\vec q - \vec q_r\right),
\end{equation}
where each entry of $\vec x$ has an analytical form. Note that all the components of $\vec x$ are expected to be small (in norm), since the points $\vec q$ and $\vec q_r$ are very close (note that $\nu$ will be chosen such that the factor $\nu/(1-\nu)$ is not too large). In fact, it can be proven that the point $\vec x$ is local, using the following result:

\emph{Lemma 1.} A correlation point $\vec x$ is local if $\sum_{i,j}\left|x_{ij}\right|<1$.

The proof, as well as more details on this analysis, can be found in Appendix C.

Next, setting $\nu=999/1000$ we obtain the bound
\begin{equation}
\label{91per100}
\frac{\nu}{1-\nu}\sum_{x,y=1}^{m}\left|q(x,y)-q_r(x,y)\right| < 1,
\end{equation}
implying that the point $\vec q$ is local via the above lemma.

To summarize, we obtain that the following lower bound for the critical visibility

\begin{equation}
v_c  \geq \eta^2\nu v_0 \simeq 0.6829
\end{equation}
where $\eta=\cos^2(\pi/50)$ is the shrinking factor of the polyhedron.
We provide a Mathematica file which gives the points $\vec q$ and $\vec q_r$ and checks the validity of condition~(\ref{91per100}), as well as a file containing all the data for the proof. This material is available online \cite{webpage_tamas}.

\section{LHV model for POVMs}

We also provide a better LHV model for Werner states considering arbitrary local measurements, i.e. moving from projective measurements to general POVMs. Specifically, we give a model for  $v_2 = 2v_1/3 \simeq 0.4553$. This improves on a previous model of Barrett \cite{barrett02} that reached $v = 5/12\simeq 0.4167$.

The construction of the model is based on the following argument. Essentially, any noisy qubit POVM can be expressed as a convex combination involving only projective qubit measurements, given the amount of white noise is above a certain threshold $\mu$. Therefore, if the statistics of certain Werner states $\rho(v) $ admit a LHV model, then so do the statistics of noisy POVMs (given the amount of noise is above the threshold $\mu$). Again, this follows from the linearity of the trace rule. In turn this implies that the statistics of arbitrary (noiseless) POVMs on the slightly more noisy Werner state $\rho(\mu^2 v) $ is local.

More formally, we can make the following statement.

\emph{Lemma 2.}
Any noisy qubit POVM $M(\mu)$ with elements $\{M_i(\mu)\}_{i=1,\ldots,4}$ proportional to rank-1 projectors for $\mu = \sqrt\frac{2}{3} - \epsilon$ can be written as a convex sum of rank-1 projectors, where $\epsilon$ may be arbitrary close to zero.

The proof is deferred to Appendix D. The above value $\mu = \sqrt{2/3}$ along with our lower bound $v_1$ for $v_c$ implies the lower bound of the visibility
\begin{equation}
\mu^2 v_1 = \frac{2}{3}v_1 = \frac{2 \cdot 999 \cdot 689}{3 \cdot 10^6} \cos^4\left(\frac{\pi}{50}\right) \simeq 0.4553
\end{equation} for a LHV model for POVMs.
We also refer to an independent related work \cite{oszmaniec16}, in particular for an alternative proof of Lemma 2.

\section{Discussion}

We have presented better LHV models for Werner states, as well as a new upper bound on the Grothendieck constant of order 3. The methods we develop provide analytical bounds, which will converge to the exact value of $v_c$ (and $K_G(3)$) using increased computational power.

Clearly, these methods can be applied to construct LHV models for other classes of entangled states, in particular in higher Hilbert space dimensions. It would also be interesting to adapt the present technique to construct local hidden state models, a specific class of LHV models relevant in the context of quantum steering \cite{wiseman07}. Finally, these methods could also be used to obtain bounds on other Grothendieck constants. While this looks computationally challenging at first sight, taking advantage of symmetry arguments could lead to progress.

\emph{Note added.} In a related work, the authors of Ref. \cite{oszmaniec16} also presented a better LHV model for Werner states for POVMs, achieving a visibility similar to ours.

\emph{Acknowledgements.} The authors thank Denis Rosset and Leonardo Guerini for discussions. F.H., M.T.Q. and N.B acknowledge financial support from the Swiss National Science Foundation (Starting grant DIAQ and QSIT); M.T.Q. acknowledges support from Japan Society for the Promotion of Science (JSPS) by KAKENHI grant No. 16F16769; T.V. from the Hungarian National Research Fund OTKA (K111734); M.N. from the FQXi grant "Towards and almost quantum physical theory".




\bibliographystyle{linksen}
\bibliography{mtqbib.bib}


\section{Appendix A. Description of the heuristic oracle}
\label{appA}

The oracle returning $\vec l_i$ in maximizing the overlap
\begin{equation}
\label{obj}
S=(\vec q-\vec q_i)\cdot\vec l
\end{equation}
over all $\vec l\in\mathcal{L}$ is a heuristic one. This problem corresponds to Step~2 in the algorithm in Sec. 2 of the main text. It is first noted that it is enough to maximize over all vertices $\vec D_{\lambda}$ of the set $\mathcal{L}$, since $\mathcal{L}$ is a polytope. The objective $S$ in Eq.~(\ref{obj}) for a given strategy $\lambda=(a_1,a_2,\ldots,a_m,b_1,b_2,\ldots,b_m)$ (and corresponding vertex $\vec D_{\lambda}$) is as follows:
\begin{equation}
\label{objab}
S_{\lambda}=\sum_{x,y=1}^m{(q(x,y)- q_i(x,y))a_xb_y}.
\end{equation}
The number of vertices (and different $\lambda$ strategies) is $2^{2m}$, hence evaluating $S_{\lambda}$ for all $\lambda$ and picking the biggest one is clearly not tractable in our range of $m>600$. Therefore, instead of a brute force computation we have to resort to a heuristic method. Note that a heuristic method still suffices, since the intuition behind $\vec l_i$ is that it gives a direction for $\vec q_i$ to move towards a better point.

The iterative algorithm is as follows.

\begin{enumerate}
\item Choose randomly assignments $\{a_x\in\pm 1\}_x$ for the deterministic strategy.
\item Fixing $\{a_x\}_x$, maximize $S$ in function of $b_y$. This amounts to setting $b_y=+1$ if $\sum_x{\left(q(x,y)-q_i(x,y)\right)a_x}>0$, otherwise to setting $b_y=-1$ for all $y=1,\ldots,m$.
\item Fixing $\{b_y\}_y$, maximize $S$ in function of $a_x$. This amounts to setting $a_x=+1$ if $\sum_y{\left(q(x,y)-q_i(x,y)\right)b_y}>0$, otherwise to setting $a_x=-1$ for all $x=1,\ldots,m$.
\item Go back to step 2 until convergence of $S$ is reached.
\end{enumerate}

Note that in each iteration step the value of the objective $S$ is guaranteed not to decrease. However, the algorithm may easily get stuck in a non-optimal $S$. To make the iterative procedure more efficient, we run it several times (in practice we chose to run it 100 times) starting from different random seeds, and pick the solution $\lambda$ and corresponding vertex $\vec D_{\lambda}$ with the biggest value of $S$.

\section{Appendix B. A family of polyhedra}

The polyhedra are parameterized by an integer $n$. In case of $n$ odd, the vertices $\hat{u}(i_1,i_2)$ are given by
\begin{equation}
\hat{u}(i_1,i_2) = \left(\cos\frac{i_1\pi}{n}\cos\frac{i_2\pi}{n},\sin\frac{i_1\pi}{n}\cos\frac{i_2\pi}{n},\sin\frac{i_2\pi}{n}\right),
\end{equation}
where $i_1,i_2\in\{0,1,\ldots,n-1\}$ plus their antipodal points. This amounts to $m=n^2$ measurement settings. In case of $n$ even, the number of vertices up to inversion (i.e. the number of settings) are $m=n^2-n+1$ due to possible redundancy of some of the vertices.

It is easy to derive an analytical expression for the shrinking factor associated with this polyhedron, which is given by $\eta = \cos^2(\pi/2n)$. Indeed, let us fix $i_1$, in which case the vertices $\pm\hat{u}(i_1,i_2)$, $i_2=(0,\ldots,n-1)$ define a regular $2n$-gon in the two-dimensional plane with a planar shrinking factor of $\eta_{2d}=\cos(\pi/2n)$. Any point $\hat{u}$ on the unit sphere can be written as a convex combination of two neighboring planes defined by some $i_1\in\{0,1,\ldots,n-1\}$ and $i_1+1 \pmod{n}$. In the worst case, the point $\hat{u}$ lies just midway between these two planes, in which case we get the lower bound of $\eta = \eta_{2d}\cos(\pi/2n)=\cos^2(\pi/2n)$ on the 3-dimensional shrinking factor.

\section{Appendix C. Going from numerical to exact precision}
\label{appC}

The algorithm described in the main text does not provide us precisely the point $\vec q$ but only a point $\vec q_{\epsilon}$, which is very close to $\vec q$, say,
\begin{equation}
\label{epsi}
\|\vec q - \vec q_{\epsilon}\|\le \epsilon.
\end{equation}
Note that $\vec q_{\epsilon} = \sum_{\lambda} w_{\lambda} \vec D_{\lambda}$, where the (positive) weights $w_{\lambda}$ coming from the algorithm are given in double precision format, whereas $\vec D_{\lambda}$ are deterministic strategies with $\pm 1$ entries.

In order to provide an analytical proof, we first transform the positive floating-point numbers $w_{\lambda}$ to positive rationals $w^r_{\lambda}$. To this end, we use the truncation
\begin{equation}
w^r_{\lambda} = \frac{\lfloor 10^{k}w_{\lambda}\rfloor}{10^{k}},
\end{equation}
where $k$ denotes the number of figures kept behind the decimal point. Then we renormalize and obtain the rational point
\begin{equation}
\vec q_r=\frac{\sum_{\lambda} w^r_{\lambda} \vec D_{\lambda}}{\sum_{\lambda} w^r_{\lambda}}
\end{equation}
which is local by construction. Then we can write
\begin{equation}
\label{vecq}
\vec q = \vec q_r + \vec q_\textrm{junk},
\end{equation}
where $\vec q_\textrm{junk}$ takes care of the (small) difference between the analytical points $\vec q$ and $\vec q_r$.
Let us now slightly shrink the point $\vec q$ towards the centre of the local polytope by rescaling $\vec q$ with a factor $\nu$ smaller than 1:
\begin{equation}
\label{vecpq}
\nu\vec q = \nu\vec q_r + \nu\vec q_\textrm{junk} = \nu\vec q_r + (1-\nu)\frac{\nu\vec q_\textrm{junk}}{1-\nu}.
\end{equation}
From the right-hand side expression, it is clear that $\nu\vec q$ is provenly local if the point $\vec x = \nu\vec q_\textrm{junk}/(1-\nu)$ is local as well. By rearranging (\ref{vecpq}), we have
\begin{equation}
\vec x = \frac{\nu}{1-\nu}\left(\vec q - \vec q_r\right),
\end{equation}
where each entry of $\vec x$ has an analytical form. Moreover, due to equation~(\ref{epsi}), the entries of $\vec x$ are typically small in case of small $\epsilon$, and $\nu$ not extremely close to 1. This suggests an easy test to decide if $\vec x$ is local. Namely,

\emph{Lemma 1.} A correlation point $\vec z$ is local if $\sum_{i,j}\left|z_{ij}\right|<1$.

\begin{proof} The proof is based on an explicit decomposition
\begin{equation}
\vec z =\sum_{i,j}\left|z_{ij}\right|\sign(z_{ij})\vec E_{i,j}+\left(1-\sum_{i,j}\left|z_{ij}\right|\right) \vec E_0
\end{equation}
in terms of local points $\pm\vec E_{i,j}$ and $\vec E_0$, where all entries of the point $\vec E_{i,j}$ are zero but entry $(i,j)$ where it takes up $+1$. On the other hand, $\vec E_0$ stands for the distribution with all entries zero. From the positivity of the weights in the above decomposition, it follows that $\vec z$ admits a LHV model if $\sum_{i,j}\left|z_{ij}\right|<1$ as claimed in Lemma~2. Note also that $\sum_{i,j}\left|z_{ij}\right|<1$ entails that all entries are bounded by $\pm 1$, hence such a $\vec z$ is a valid correlation point by definition.
\end{proof}

\section{Appendix D. Decomposing noisy qubit POVMs in terms of projectors}
\label{appD}

An extremal POVM $M$ for qubits can be characterized as follows. It has no more than four elements $\{M_i\}$, $i=1,2,3,4$ such that each element $M_i$ is proportional to a rank 1 projector \cite{dariano05}. Let us define the vector of Pauli matrices $\vec\sigma=(\sigma_x,\sigma_y,\sigma_z)$ and write the POVM elements in the form
\begin{equation}
M_i = a_i \openone + \vec a_i\cdot\vec\sigma,
\end{equation}
where $a_i=|\vec a_i|$, $\sum_i a_i = 1$, and $\sum_i \vec a_i = 0$.

\noindent Similarly, we define the elements of a noisy POVM as follows
\begin{equation}
M_i(\mu) = a_i \openone + \mu\vec a_i\cdot\vec\sigma,
\end{equation}
where $0\le \mu\le 1$. Note that $\mu=1$ corresponds to the noiseless case. Also an arbitrary qubit rank 1 projector $P$ can be described with the following two elements
\begin{eqnarray}
P_1 &=& (1/2) \openone + (1/2)\vec b\cdot\vec\sigma,\nonumber\\
P_2 &=& (1/2) \openone - (1/2)\vec b\cdot\vec\sigma,
\end{eqnarray}
where $\vec b$ is a unit vector.

With these definitions, we state our Lemma~1 in the main text:

Any noisy qubit POVM $M(\mu)$ with elements $\{M_i(\mu)\}_{i=1,\ldots,4}$ proportional with rank 1 projectors for $\mu = \sqrt\frac{2}{3} - \epsilon$ can be written as a convex sum of rank 1 projectors $P^{(k)}$, where $\epsilon$ may be arbitrary close to zero and $k$ may run up to infinity.

In order to prove it, we start with the following lemma.

\emph{Lemma 3.}
Any noisy POVM $M(\mu)$ with elements $\{M_i(\mu)\}$, $i=1,2,3,4$ for $\mu = \sqrt\frac{2}{3} - \epsilon$ can be expressed in terms of a projector $P$ and another noisy POVM $M'(\mu)$ with the same factor $\mu$ as follows
\begin{equation}
\label{decomp}
M(\mu) = p P + (1-p)M'(\mu),
\end{equation}
where $p$ is strictly larger than zero if $\epsilon>0$.

Note that the above lemma already provides us a constructive method to prove Lemma 1: We start with $M(\mu)$ and use Lemma~3 to decompose it as
\begin{equation}
M(\mu) = p_1 P^{(1)} + (1-p_1)M^{(1)}(\mu)
\end{equation}
with $p_1>0$. Then we run the protocol again starting with $M^{(1)}(\mu)$
\begin{equation}
M^{(1)}(\mu) = p_2 P^{(2)} + (1-p_2)M^{(2)}(\mu)
\end{equation}
to get the decomposition
\begin{equation}
M(\mu) = p_1 P^{(1)} + (1-p_1)p_2 P^{(2)} + (1-p_1)(1-p_2)M^{(2)}(\mu).
\end{equation}
Running this iterative procedure up to $n$ times, we get a decomposition of $M(\mu)$ in terms of $n$ projectors $P^{(k)}$, $k=1,2,\ldots,n$ and a POVM $M^{(n)}(\mu)$ with an overall weight of $\prod_{k=1}^{n} (1-p_k)$ of this POVM. If we were demanding each $p_k$ to be maximal at each step $k$, then, as $n$ goes to infinity, this weight must go to zero. For, if it did not, then there would be a subsequence of $(M^{(k)}(\mu))_k$ converging to some POVM $\bar{M}(\mu)$ with the property that it cannot be decomposed further, a contradiction. Therefore one arrives at a decomposition of $M(\mu)$ only in terms of projectors.

We are now left with a proof of Lemma~3. To this end, we state another lemma.

\emph{Lemma 4.}
Given four nonzero vectors $\vec a_i$, $i=1,2,3,4$ in the three-dimensional Euclidean space such that they sum up to zero. Then the relation
\begin{equation}
\label{ijpair}
\frac{\vec a_i\cdot\vec a_j}{a_i a_j}\le -\frac{1}{3}
\end{equation}
holds true for at least one pair (say, $i$ and $j$), where we defined $a_i=|\vec a_i|$. In other words, one can always pick two vectors for which the angle $\theta_{ij}$ between them is at least $\theta_{ij}=\arccos(-1/3)$. Note the special case of the vertices of the regular tetrahedron for which each angle $\theta_{ij}$ between the vectors formed by the vertices is $\arccos(-1/3)$.

\begin{proof}
The proof is by contradiction. Suppose that the lemma is not true and we can pick vectors $\vec a_i$ such that
\begin{eqnarray}
\vec a_1\cdot\vec a_4&>& -a_1 a_4/3\nonumber\\
\vec a_2\cdot\vec a_4&>& -a_2 a_4/3\nonumber\\
\vec a_3\cdot\vec a_4&>& -a_3 a_4/3.
\end{eqnarray}
Then summing up the above three equations, and plugging $\vec a_4 = -\vec a_1-\vec a_2-\vec a_3$, we get $(a_1+a_2+a_3)>3a_4$. If we choose for instance the ordering $a_4\ge a_3\ge a_2\ge a_1$, the above relation is clearly not true.
\end{proof}

With the above tools, we are ready to prove Lemma~3 which in turn proves Lemma~1. To this end we write out equation~(\ref{decomp}) for each POVM element and we assume w.l.o.g. that the ($i,j$) pair satisfying equation~(\ref{ijpair}) is given by $(i,j)=(1,2)$. Then we have
\begin{eqnarray}
\label{scalars}
a_1 &=& (p/2) + (1-p)c_1\nonumber\\
a_2 &=& (p/2) + (1-p)c_2\nonumber\\
a_3 &=& (1-p)c_3\nonumber\\
a_4 &=& (1-p)c_4
\end{eqnarray}
for the scalar terms and
\begin{eqnarray}
\label{vectors}
\mu\vec a_1 &=& (p/2)\vec b + (1-p)\mu\vec c_1\nonumber\\
\mu\vec a_2 &=& (-p/2)\vec b + (1-p)\mu\vec c_2\nonumber\\
\mu\vec a_3 &=& (1-p)\mu\vec c_3\nonumber\\
\mu\vec a_4 &=& (1-p)\mu\vec c_4
\end{eqnarray}
for the vectors, where the other POVM $M'(\mu)$ in (\ref{decomp}) is defined by the elements $M'_i(\mu) = c_i + \mu\vec c_i\cdot\vec\sigma$. Let us denote the angle between $\vec a_1$ and $\vec a_2$ by $\theta_{12}$, which we write in terms of two positive angles $\theta_1$, $\theta_2$ as follows $\theta_{12}=\theta_1+\theta_2$ which are yet to be determined.
After an appropriate rotation of the coordinate system, we can use the following parametrization
\begin{eqnarray}
\label{subset}
\vec a_1 &=& \sin\theta_1 e_x + \cos\theta_1 e_z \nonumber\\
\vec a_2 &=& -\sin\theta_2 e_x + \cos\theta_2 e_z \nonumber\\
\vec b &=& e_x
\end{eqnarray}
and we can also assume w.l.o.g. that $a_1\ge a_2$.

Using the first equation of (\ref{vectors}), we separate the $e_x$ and $e_z$ terms which result in two equations
\begin{eqnarray}
\label{couple}
\mu a_1\sin\theta_1 &=& (p/2) + (1-p)\eta c_{1x} \nonumber\\
\mu a_1\cos\theta_1 &=& (1-p)\mu c_{1z},
\end{eqnarray}
where we defined $\vec c_i = c_{ix} e_x + c_{iz} e_z$ for $i=1,2$. Combining (\ref{couple}) with the first equations of (\ref{scalars}) and (\ref{vectors}), $p$ can be expressed as follows:
\begin{equation}
\label{p1}
p = \frac{4a_1\mu}{1-\mu^2}(\sin\theta_1-\mu).
\end{equation}
Similarly, from the second equations of (\ref{scalars}) and (\ref{vectors}), we arrive at
\begin{equation}
\label{p2}
p = \frac{4a_2\mu}{1-\mu^2}(\sin\theta_2-\mu).
\end{equation}

Given $\theta_{12}$, $a_1\ge a_2>0$ (which define $M(\mu)$), our goal is to find a $p$ strictly greater than 0. Note that due to Lemma~4 we can also assume that $\theta_{12}\ge\arccos(-1/3)$. Note also that $a_1=0$ entails $a_2=0$, which means that $M(\mu)$ is already a projector and the proof can be finished. On the other hand, if $a_2=0$ we go back to the case of a three-outcome POVM which has to be treated similarly to the general four-outcome situation and will be discussed later.

Let us split the general case $a_1\ge a_2$ into $a_1=a_2$ and $a_1>a_2$.

We start with $a=a_1=a_2$. Here we take $\theta_1=\theta_2=\theta_{12}/2$ and the pair of equations~(\ref{p1},\ref{p2}) become a single equation for $p$:
\begin{equation}
\label{psimple}
p = \frac{4a\mu}{1-\mu^2}(\sin(\theta_{12}/2)-\mu).
\end{equation}
Since we discuss the noisy situation $0<\mu<1$, the factor $\frac{4a\mu}{1-\mu^2}$ is positive and we get the condition $\sin(\theta_{12}/2)>\mu$ to get an $\mu$ for which $p$ is strictly larger than zero. The worst case scenario is defined by $\theta_{12}=\arccos(-1/3)$ which gives the critical value $\mu_{crit}=\sin(\arccos(-1/3)/2)=\sqrt{2/3}$ below which $p$ is strictly larger than zero.

We now discuss the case $a_1>a_2$. From the pair of equations~(\ref{p1},\ref{p2}) it is clear that if we want to satisfy both with a single value of $p$, it is required to have $\sin\theta_2>\sin\theta_1$. Recalling that $\theta_1+\theta_2=\theta_{12}\ge\arccos(-1/3)$, where $\theta_1$  and $\theta_2$ are positive, we get the lower bound $\theta_2\ge\arccos(-1/3)/2$ which entails the bound $\sin\theta_{2}\ge \sqrt{2/3}$.

Altogether we obtain the result that in case of $a_1\ge a_2>0$, $p$ is strictly larger than zero whenever $\mu_{crit} = \sqrt{2/3}-\epsilon$, where $\epsilon$ can be arbitrary small.

Let us now come back to the situation when $a_2=0$. In that case, we get the very same equations for $p$ as in Eqs.~(\ref{p1},\ref{p2}) with the only exception that instead of Lemma~4, we have

\emph{Lemma 5.}
Given three nonzero vectors $\vec a_i$, $i=1,2,3$ in the three-dimensional Euclidean space such that they sum up to zero. Then the relation
\begin{equation}
\label{ijpair3}
\frac{\vec a_i\cdot\vec a_j}{a_i a_j}\le -\frac{1}{2}
\end{equation}
holds true for at least one pair (say, $i$ and $j$), where we defined $a_i=|\vec a_i|$.

The proof is analogous to the one of Lemma~4.
However, the new condition in (\ref{ijpair3}) implies the even larger bound of $\mu_{crit}=\sqrt{3}/2$. Hence, the case $a_2=0$ is solved as well. As a side result, we also have a theorem for 3-outcome POVMs:

\emph{Lemma 6.}
Any noisy three-outcome POVM $M(\mu)$ with elements $\{M_i(\mu)\}$, $i=1,2,3$ for $\mu = \sqrt{3}/2 - \epsilon$ can be written as a convex sum of projectors $P^{(k)}$, where $\epsilon$ may go arbitrary close to zero and $k$ may run up to infinity.

\end{document}